\documentclass[conference]{IEEEtran}
\usepackage{cite}
\usepackage{amsmath,amssymb,amsfonts}
\usepackage{algorithmic}
\usepackage{graphicx}
\usepackage{textcomp}
\usepackage{xcolor}
\usepackage{url}
\def\BibTeX{{\rm B\kern-.05em{\sc i\kern-.025em b}\kern-.08em
    T\kern-.1667em\lower.7ex\hbox{E}\kern-.125emX}}
\begin{document}

\title{Secure Federated XGBoost with CUDA-accelerated Homomorphic Encryption via NVIDIA FLARE
}

\author{
\IEEEauthorblockN{Ziyue Xu, Yuan-Ting Hsieh, Zhihong Zhang, Holger R. Roth, Chester Chen, Yan Cheng, and Andrew Feng}
\IEEEauthorblockA{\textit{Nvidia Corp.}, USA}
}

\maketitle

\begin{abstract}
Federated learning (FL) enables collaborative model training across decentralized datasets. NVIDIA FLARE's Federated XGBoost extends the popular XGBoost algorithm to both vertical and horizontal federated settings, facilitating joint model development without direct data sharing. However, the initial implementation assumed mutual trust over the sharing of intermediate gradient statistics produced by the XGBoost algorithm, leaving potential vulnerabilities to honest-but-curious adversaries. This work introduces ``Secure Federated XGBoost'', an efficient solution to mitigate these risks. We implement secure federated algorithms for both vertical and horizontal scenarios, addressing diverse data security patterns. To secure the messages, we leverage homomorphic encryption (HE) to protect sensitive information during training. A novel plugin and processor interface seamlessly integrates HE into the Federated XGBoost pipeline, enabling secure aggregation over ciphertexts. We present both CPU-based and CUDA-accelerated HE plugins, demonstrating significant performance gains. Notably, our CUDA-accelerated HE implementation achieves up to 30x speedups in vertical Federated XGBoost compared to existing third-party solutions. By securing critical computation steps and encrypting sensitive assets, Secure Federated XGBoost provides robust data privacy guarantees, reinforcing the fundamental benefits of federated learning while maintaining high performance.
\end{abstract}


\section{Introduction}
XGBoost~\cite{xgboost} is a machine learning algorithm widely used for tabular data modeling. It is highly effective and scalable for common regression and classification tasks. Building on the principles of gradient boosting, it combines the predictions of multiple sequentially-learnt sub-models, typically decision trees, to produce a robust overall model. DMLC XGBoost\footnote{\url{https://github.com/dmlc/xgboost/}} provides a scalable solution for large datasets and intricate data structures due to its optimized implementation and advanced capabilities, including regularization, parallel computing, and robust handling of missing values. Its efficiency and adaptability have contributed to its widespread use in data science competitions and real-world applications across multiple industries.

To expand the XGBoost model from single-site learning to multi-site collaborative training~\cite{fl_general}, NVIDIA has developed ``Federated XGBoost'', an XGBoost plugin for federation learning (FL). It covers vertical collaboration settings to jointly train XGBoost models across decentralized data sources. It was first released in XGBoost 1.7.0, enabling multiple institutions to jointly train XGBoost models without the need to centralize the data; then, it was further extended in XGBoost 2.0.0 release to support vertical FL. 

To fully support an industry-level federated XGBoost pipeline, Federated XGBoost needs to be integrated with an FL framework. NVIDIA Federated Learning Application Runtime Environment (FLARE\footnote{\url{https://github.com/NVIDIA/NVFlare}})~\cite{nvflare_main}, a domain-agnostic, open-source, and extensible SDK for FL, has enhanced the real-world FL experience by introducing capabilities to handle communication challenges. This includes multiple concurrent training jobs, and potential job disruptions due to network conditions. Since 2023, NVIDIA FLARE has introduced built-in integration with Federated XGBoost features~\cite{fedxgb_techblog}:  horizontal histogram-based and tree-based XGBoost, as well as vertical XGBoost. We have also added support for Private Set Intersection (PSI) for sample alignment as a preprocessing step for vertical training. 

\begin{figure}[bthp]
\centering
\includegraphics[width=0.5\textwidth]{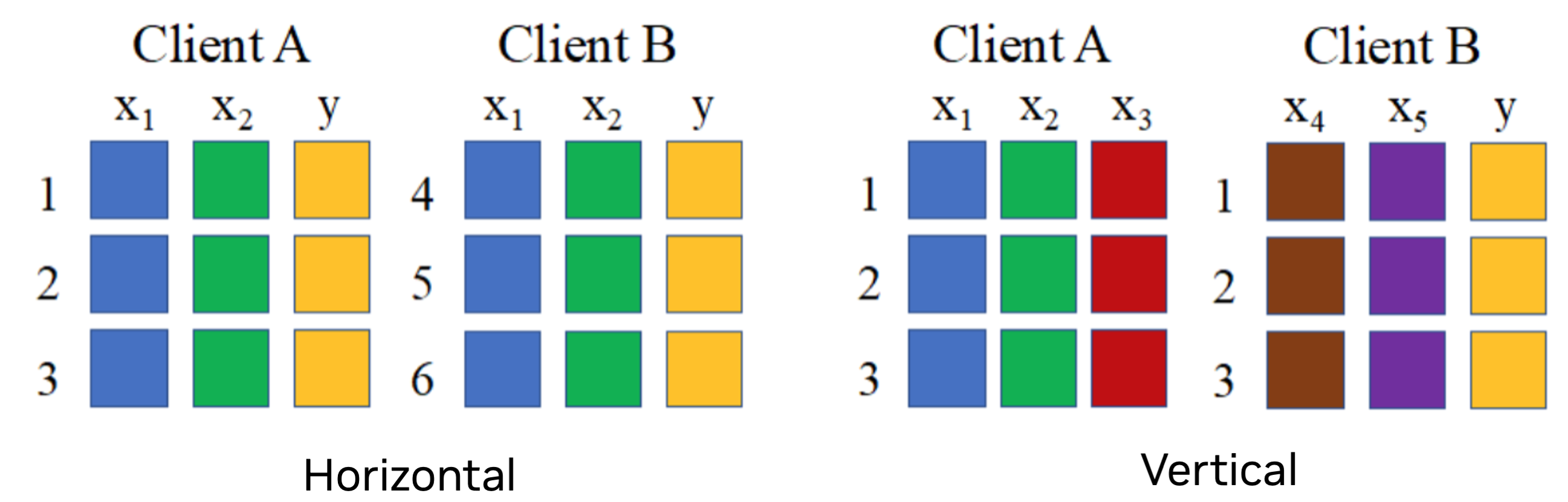}
\caption{Partitioning of features $x$ and labels $y$ in horizontal and vertical federated learning setups.}
\label{fig:hori_vert}
\end{figure}

With this integration, reliable Federated XGBoost with comprehensive experiment tracking and other support features helps to streamline the practical usage of decentralized XGBoost training. However, there are still concerns unaddressed, especially with regard to data privacy and security. The previous federated XGBoost is built with the assumption of full mutual trust, indicating that no party has the intention to learn more information beyond model training. In practice, however, honest-but-curious is a more realistic setting for federated collaborations. Under this setting, participants would want to learn additional information based on the data being exchanged, including but not limited to recovering the label information from the sample-wise gradients, and evaluating the feature characteristics according to the gradient histograms.

In this work, NVIDIA FLARE and XGBoost expand the scope of Federated XGBoost by further securing these potential information concerns. Specifically:
\begin{itemize}
    \item The secure federated algorithms, both horizontal and vertical, are implemented and added to the federated schemes supported by the XGBoost library, addressing data security patterns under different assumptions. This includes both training and inference.
    \item Homomorphic encryption (HE) features are added to the secure federated XGBoost pipelines using a plugin and processor interface system designed to robustly and effectively bridge the two libraries: the computation by XGBoost, and communication by NVIDIA Flare with proper encryption, aggregation, and decryption processes in between. 
    \item HE plugins are developed, both CPU-based and CUDA-accelerated, providing versatile adaptation depending on hardware and efficiency requirements. The CUDA plugin is shown to be much faster than current third-party solutions.
\end{itemize}

With the help of HE, key federated computation steps are performed over cipher-texts, and the relevant assets (gradients and partial histograms) are encrypted and will not be learned by other parties during computation. This gives users assurance of their data security, which is one of the fundamental benefits of federated learning. Further, CUDA-accelerated HE with Federated XGBoost adds security protection for data privacy and delivers up to 30x speedups for vertical XGBoost compared to third-party solutions.

\section{Collaboration Modes and Secure Patterns}
Collaboration modes can be viewed from the perspectives of both data distribution and algorithmic process. Depending on their combinations, we can have various secure patterns.

\subsection{Collaboration Modes}
\subsubsection{Data Split}
Considering the data split and distribution, there are mainly two collaboration modes: horizontal and vertical. Under the horizontal setting, each participant holds all features and label information, but only for part of the whole population. While under the vertical setting, each party holds part of the features for the entire population, and only one party holds the label. The label-owner is referred to as the active party, while all other parties are passive parties. 

Fig.~\ref{fig:hori_vert} illustrates a simple case for horizontal and vertical collaborations:
\begin{itemize}
    \item In horizontal case, each participant has access to the same features (columns - ``$x_1$ $x_2$'') and label (``y'') of different data samples (rows - 1/2/3 for Client A v.s. 4/5/6 for Client B). In this case, everyone holds equal status as ``label owner''.
    \item In vertical case, each client has access to different features (columns - ``$x_1$ $x_2$ $x_3$'' for Client A v.s. ``$x_4$ $x_5$'' for Client B) of the same data samples (rows - 1/2/3). We assume that only one is the ``active party'', i.e. ``label owner'' - Client B owning label ``y''.
\end{itemize}

\subsubsection{Algorithm}
From an algorithmic perspective, we can also have two collaboration modes: histogram-based and tree-based. 

The histogram-based collaboration leverages the federated learning support in XGBoost: allowing the existing distributed XGBoost training algorithm to operate in a federated manner, with the federated clients acting as the distinct workers in the distributed XGBoost algorithm. In this scenario, individual federated participants share and aggregate gradient information about their respective portions of the training data, as required to optimize tree node splitting when building the successive boosted trees. Virtually, such federated learning process is identical to that of a distributed XGBoost model learning.

The shared information is in the form of quantile sketches of feature values as well as corresponding sample gradient and sample Hessian histograms (``Local G/H''), based on which the global information can be computed (``Global G/H''). Under federated histogram-based collaboration, information of precisely the same structure is exchanged among the clients. The main differences are that the data is partitioned across the workers according to client data ownership, rather than being arbitrarily partitioned, and all communication is via an aggregating federated gRPC server instead of direct client-to-client communication. Histograms from different clients, in particular, are aggregated in the server and then communicated back to the clients.

Essentially, each tree relies on information from all federated clients collaboratively, and is thus similar / identical to the model built with centralized training.

\begin{figure*}[bthp]
\centering
\includegraphics[width=0.8\textwidth]{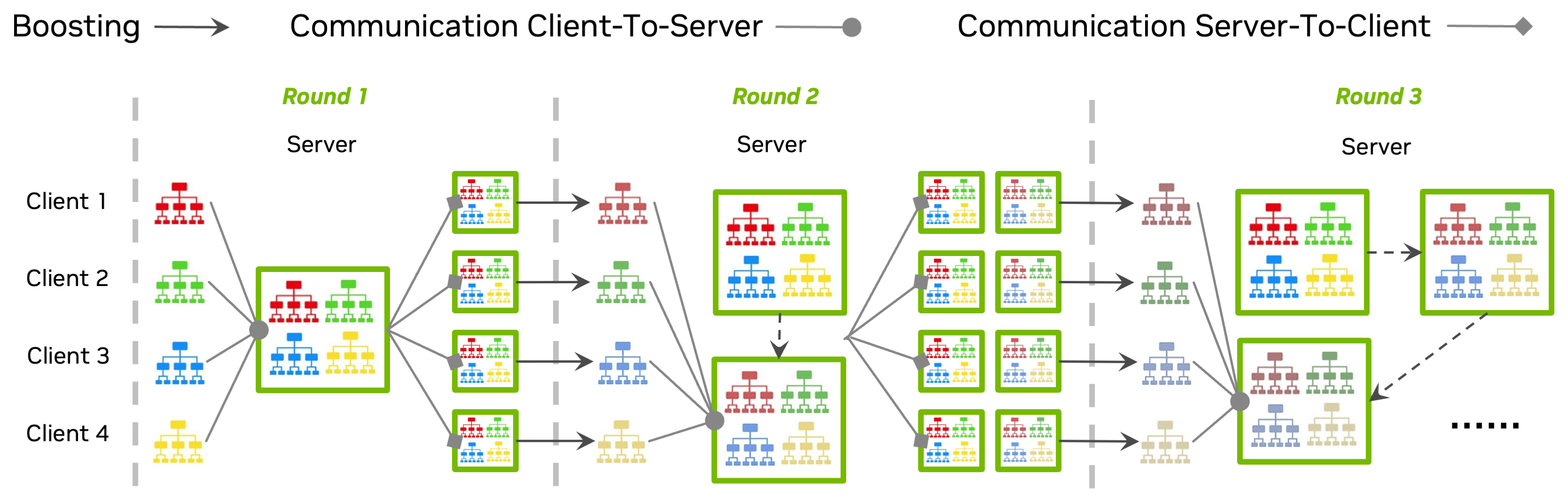}
\caption{Bagging Tree-based Federated XGBoost}
\label{fig:bagging}
\end{figure*}

In contrast, under tree-based collaboration, individual trees are independently trained on each client's local data without aggregating the global sample gradient histogram information. Trained trees are collected and passed to the server / other clients for aggregation and / or further boosting rounds. Comparing with histogram-based collaboration, the major difference is that the histogram-based methods exchange the intermediate results for tree-boosting, while tree-based methods exchange the final tree model. Thus each tree is built with global information for histogram-based methods, while with local information for tree-base methods.

Under this setting, we can further distinguish between two types of tree-based collaboration: cyclic and bagging:

For cyclic training, at each round of tree boosting, instead of relying on the whole data statistics collected from all clients, the boosting relies on only one client's local data. The resulting tree sequence is then forwarded to the next client for next round's boosting. One full ``cycle'' will be complete when all clients have been covered.

For federated XGBoost training with bagging aggregation, as illustrated in Fig.~\ref{fig:bagging}, at each round of tree boosting, all participants start from the same ``global model'', and boost a number of trees (in current example, one tree) based on their local data. The resulting trees are then sent to the server. A bagging aggregation scheme is applied to all the submitted trees to update the global model. Specifically, the global model is updated by aggregating the trees from all clients as a forest, and the global model is then broadcasted back to all clients for local prediction and further training. The XGBoost Booster API is leveraged to create in-memory Booster objects that persist across rounds to cache predictions from trees added in previous rounds and retain other data structures needed for training.

\subsection{Secure Patterns}
Under different collaboration modes, we can have various secure patterns. For tree-based collaborations, since participants under this collaboration mode exchange the boosted trees, i.e., part of the final model which will be made accessible to all, the pipeline is less likely to reveal sensitive information. Hence, when talking about data security and privacy~\cite{secureboost}, we will mainly focus on histogram-based methods, which exchange intermediate results and could have data privacy concerns.

\subsubsection{Horizontal Histogram-based}
For horizontal histogram-based XGBoost, each party holds ``equal status'' (whole feature and label for partial population), while the federated server performs aggregation, without owning any data. Each participant will submit its local gradient histograms to the server to be aggregated to a global histogram. As a global view of sample-wise gradients according to each feature, it can reveal local feature distributions. Therefore, it is undesirable that the local histograms are being learnt by others. In this case, clients have a concern of leaking information to the server, and to each other. Hence, the information to be protected is each client’s local histograms, against the server, and against each other.

\subsubsection{Vertical Histogram-based}
For vertical histogram-based XGBoost, only the active party has access to the label, making it the only one who is able to compute the sample-wise gradients needed by the algorithm. Hence, the first step of the vertical federated XGBoost is for the active party to compute the gradients and distribute the results to other parties. However, the gradient itself contains the label information that can be recovered. Since only the active party holds the label, it can be considered ``the most valuable asset'' for the whole process, and should not be accessed by passive parties. Therefore, the active party in this case is the ``major contributor'' from a model training perspective, with a concern of leaking this information to passive clients. In this case, the security protection is mainly against passive clients over the label information.

In addition, for vertical collaboration at inference time, it is less desirable for other parties to learn a specific feature's identity (e.g. whether it relates to gender, age, etc.), therefore it is important to hide such information from others, only keep it to whoever owns the feature. 

\section{Method Design}
Based on the above secure patterns, we implemented the following secure pipelines for histogram-based federated XGBoost.

\subsection{Secure Horizontal}
\begin{figure}[bthp]
\centering
\includegraphics[width=0.5\textwidth]{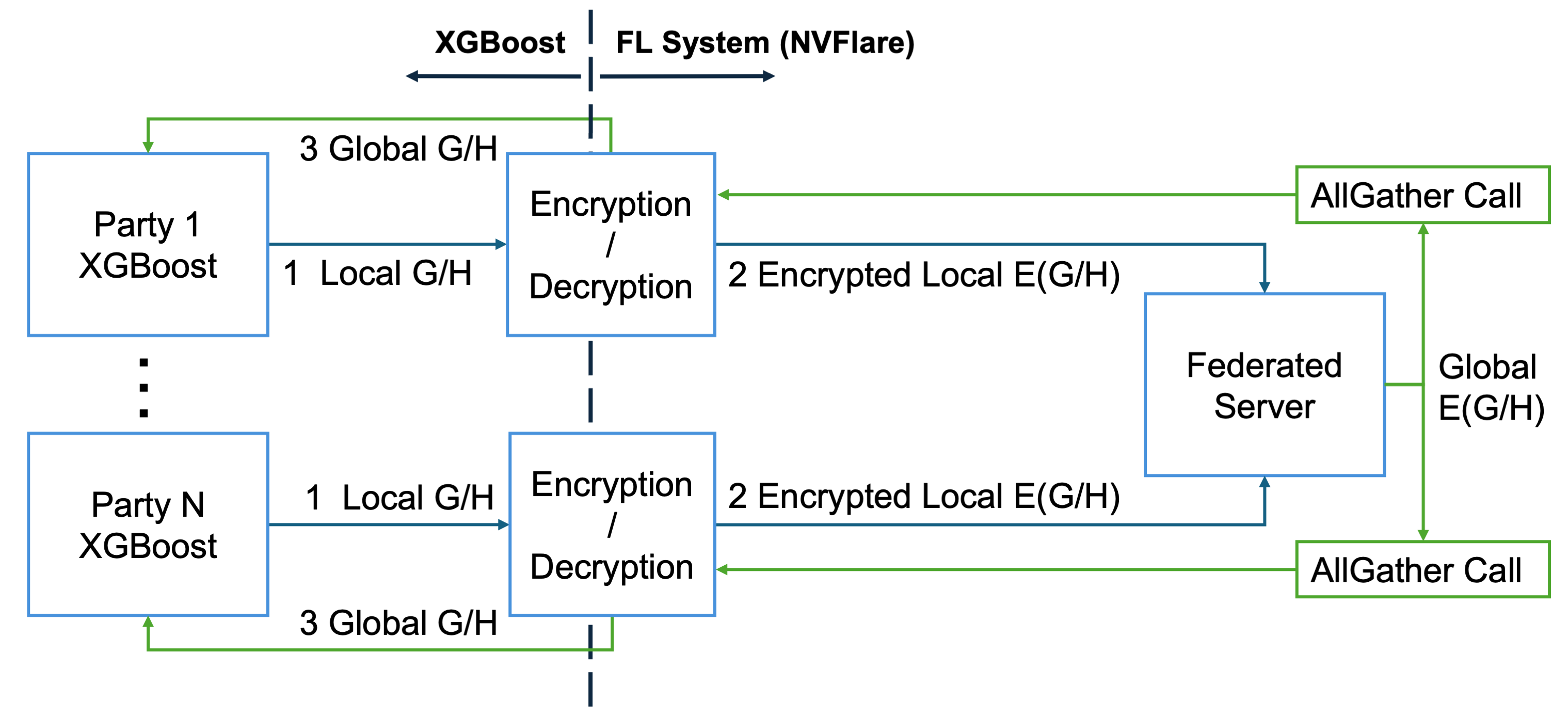}
\caption{Secure Horizontal Federated XGBoost}
\label{fig:1}
\end{figure}

As illustrated in Fig.~\ref{fig:1}, to protect the local histograms for horizontal collaboration, the histograms will be encrypted before sending to the federated server for aggregation. The aggregation will then be performed over cipher-texts and the encrypted global histograms will be returned to clients, where they will be decrypted and used for tree building. In this way, the server will have no access to the plain-text histograms, while each client will only learn the global histogram after aggregation, rather than individual local histograms.

\begin{figure}[tbph]
\centering
\includegraphics[width=0.5\textwidth]{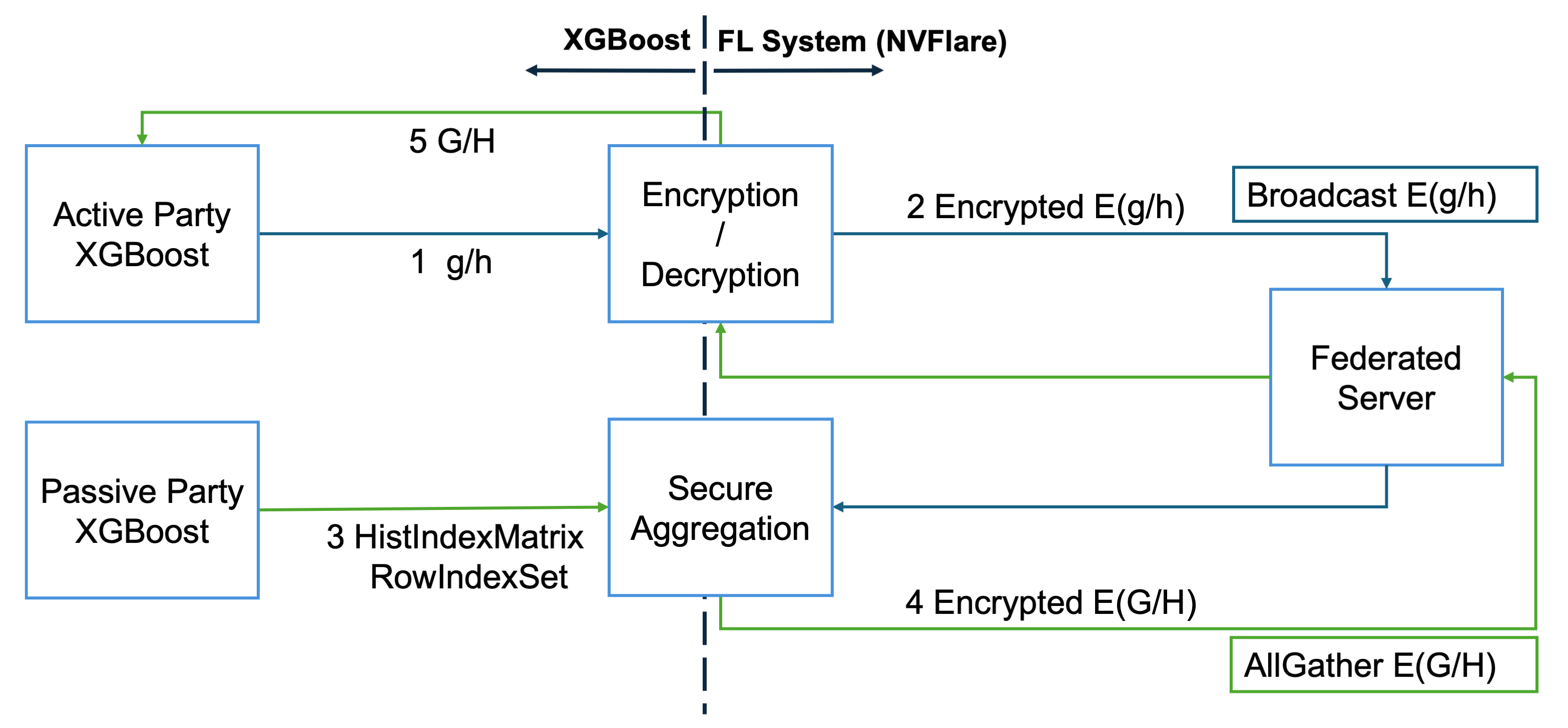}
\caption{Secure Vertical Federated XGBoost}
\label{fig:2}
\end{figure}

\subsection{Secure Vertical}
As illustrated in Fig.~\ref{fig:2}, to protect label information for vertical collaboration, at every round of XGBoost after the active party computes the gradients for each sample, the gradients will be encrypted before sending to passive parties. Upon receiving the encrypted gradients (cipher-text), they will be accumulated according to the specific feature distribution at each passive party. The resulting cumulative histograms will be returned to the active party, decrypted, and further used for tree building by the active party.

\subsection{Encryption with proper HE schemes}
With multiple libraries covering various HE schemes both with and without GPU support, it is important to properly choose the most efficient scheme for the specific needs of a particular federated XGBoost setting. Let us look at one example, assume $N=5$ number of participants, $M=200K$ total number of data samples, $J=30$ total number of features, and each feature histogram has $K=256$ slots.  Depending on the type of federated learning applications: (Vertical or Horizontal application, we will need different algorithms. 

For vertical application, the encryption target is the individual g/h numbers, and the computation is to add the encrypted numbers according to which histogram slots they fall into. As the number of g/h is the same as the sample number, for each boosting round in theory:
\begin{itemize}
 \item The total encryption needed will be $M\times2 = 400k$ (g and h), and each time encrypts a single number
 \item The total encrypted addition needed will be $(M - K)\times 2 \times J \approx 12m$
\end{itemize}

In this case, an optimal scheme choice would be Paillier~\cite{paillier} because the encryption needs to be performed over a single number. Using schemes targeting vectors would be a significant waste of space. 

For horizontal application, on the other hand, the encryption target is the local histograms G/H, and the computation is to add local histograms together to form the global histogram. For each boosting round:
\begin{itemize}
 \item The total encryption needed will be $N \times 2 = 10$ (G and H), and each time encrypts a vector of length $J \times K = 7680$
 \item The total encrypted addition needed will be $(N - 1) \times 2 = 18$
\end{itemize}

In this case, an optimal scheme choice would be CKKS~\cite{CKKS} because it is able to handle a histogram vector (with length 7680, for example) in one shot.

We provide encryption solutions both with CPU-only, and with efficient GPU acceleration. 

\subsection{Plugin Interface for Processing and Encryption}
\begin{figure*}[tbph]
\centering
\includegraphics[width=0.9\textwidth]{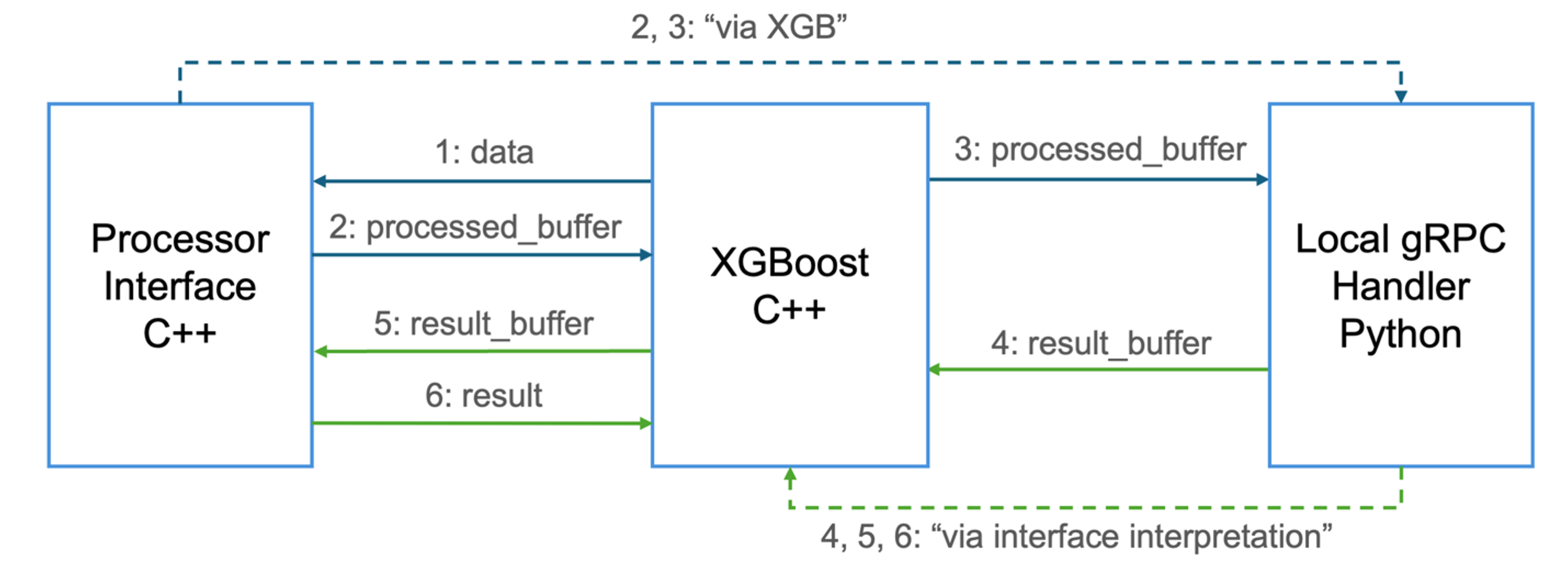}
\caption{Plugin Interface for Encryption}
\label{fig:plugin}
\end{figure*}

To couple the existing XGBoost functionality and NVFLARE federated pipeline, we designed a plugin interface for performing encryption in a versatile manner. As illustrated in Fig.~\ref{fig:plugin}, a message containing gradient information will not be directly communicated between XGBoost computer and NVFlare communicator, rather, it will first go through a processor interface as:
\begin{enumerate}
    \item Upon receiving specific MPI calls from XGBoost, each corresponding party calls interface for data processing (serialization, etc.), providing necessary information: g/h pairs, or local G/H histograms.
    \item Processor interface performs necessary processing (and encryption), and send the results back as a processed buffer.
    \item Each party then forward the message to local gRPC handler on FL system side.
    \item After FL communication involving message routing and computation, each party receives the result buffer upon MPI calls.
    \item Each FL party then sends the received buffer to processor interface for interpretation.
    \item Interface performs necessary processing (deserialization, etc.), recovers proper information, and sends the result back to XGBoost for further computation.
\end{enumerate}
This mechanism is flexible as it can be couple with any encryption plugin implementations.

\section{Experiments and Results}
With implementation of the pipeline previously described on both XGBoost and NVIDIA Flare, we tested our secure federated pipelines with a credit card fraud detection dataset. Comparing the tree models with a centralized baseline, we reached the following observations:
\begin{itemize}
 \item Vertical federated learning (non-secure) has exactly the same tree model as the centralized baseline.

 \item Vertical federated learning (secure) has the same tree structures as the centralized baseline. Furthermore, it produces different tree records at different parties because each party holds different feature subsets, and it should not learn the cut information for features owned by others.

 \item Horizontal federated learning (both secure and non-secure) have different tree models from the centralized baseline. This is due to the initial feature quantile computation, over either global data (centralized) or local data (horizontal).
\end{itemize}

\subsection{Inference Model Safety on Vertical Collaboration}
\begin{figure*}[tbph]
\centering
\includegraphics[width=0.9\textwidth]{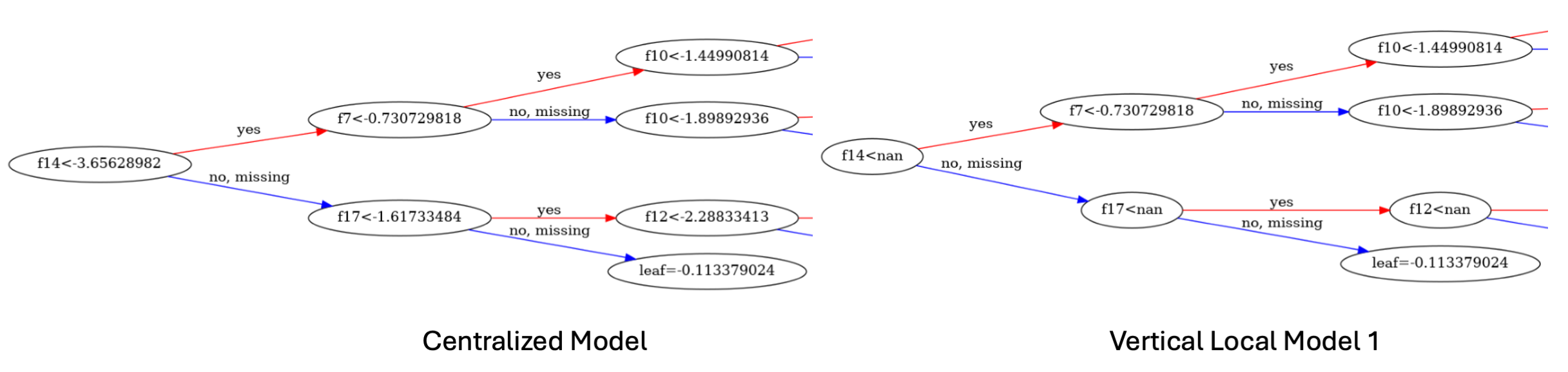}
\caption{Secure Vertical Federated XGBoost Inference with Partially-saved Model.}
\label{fig:safe_inf}
\end{figure*}

As mentioned earlier, for the final model, each feature should only reside in its owner, without being exposed to others. Therefore at each site, the local model is ``partially-saved''. As illustrated in Fig.~\ref{fig:safe_inf}, participant 1 has access to feature $\#1$ to $\#10$, while participant 2 has access to the rest of the features. Thus we can see the local model of participant 1 has `nan' values for f14, f17, and f12. Since participant 1 does not have access to these features, those nodes will not be split by this participant. At inference time, each participant will perform its own split, then the final prediction is made by combining splits from all local models, producing identical results as the centralized model. 

\subsection{Efficiency of encryption methods}
To benchmark our solutions, we conducted experiments using a diverse range of datasets with varying characteristics, including differences in size (from small to large) and feature dimensions (from few to many). These benchmarks aim to demonstrate the robustness of our algorithms and highlight significant performance improvements in terms of speed and efficiency. 

\subsubsection{Dataset and data splits}

We used three datasets, covering different data sizes and feature sizes, to illustrate their impact on the efficiency of encryption methods. The data characteristics are summarized in Table~\ref{tab:1}. The credit card fraud detection dataset is labeled as CreditCard\footnote{\url{https://www.kaggle.com/datasets/mlg-ulb/creditcardfraud}}, the Epsilon dataset\footnote{\url{https://catboost.ai/docs/en/concepts/python-reference_datasets_epsilon}} as Epsilon, and a subset of the HIGGS~\cite{higgs} dataset as HIGGS.

\begin{table}[htbp]
\caption{ Summary of the three dataset sizes for experiments, differing in the scale of both the data and the feature}
\begin{center}
\begin{tabular}{|c|c|c|c|}
\hline
 & \textbf{CreditCard} & {HIGGS} & {Epsilon} \\
\hline
Data records size & 284,807 & 6,200,000 & 400,000 \\
Feature size & 28 & 28 & 2000 \\
Training set size & 227,845 & 4,000,000 & 320,000  \\
Validation set size & 56,962 & 2,200,000 & 80,000 \\
\hline
\end{tabular}
\label{tab:1}
\end{center}
\end{table}

For vertical federated learning, we split the training dataset into two clients, with each client holding different features of the same data records (Table~\ref{tab:2}).
\begin{table}[htbp]
\caption{Summary of data for vertical federated learning}
\begin{center}
\begin{tabular}{|c|c|c|c|}
\hline
 & \textbf{CreditCard} & {HIGGS} & {Epsilon} \\
\hline
Label client & 10 & 10 & 799 \\
Non-label client & 18 & 18 & 1201 \\
\hline
\end{tabular}
\label{tab:2}
\end{center}
\end{table}

For horizontal federated learning, we split the training set into three clients evenly (Table~\ref{tab:3}).
\begin{table}[htbp]
\caption{ Summary of data for horizontal federated learning
}
\begin{center}
\begin{tabular}{|c|c|c|c|}
\hline
 & \textbf{CreditCard} & {HIGGS} & {Epsilon} \\
\hline
Client 1 & 75,948 & 1,333,333 & 106,666 \\
Client 2 & 75,948 & 1,333,333 & 106,666 \\
Client 3 & 75,948 & 1,333,334 & 106,668  \\
\hline
\end{tabular}
\label{tab:3}
\end{center}
\end{table}

\subsubsection{Computation Efficiency Comparison}
\begin{figure}[htbp]
\centering
\includegraphics[width=0.45\textwidth]{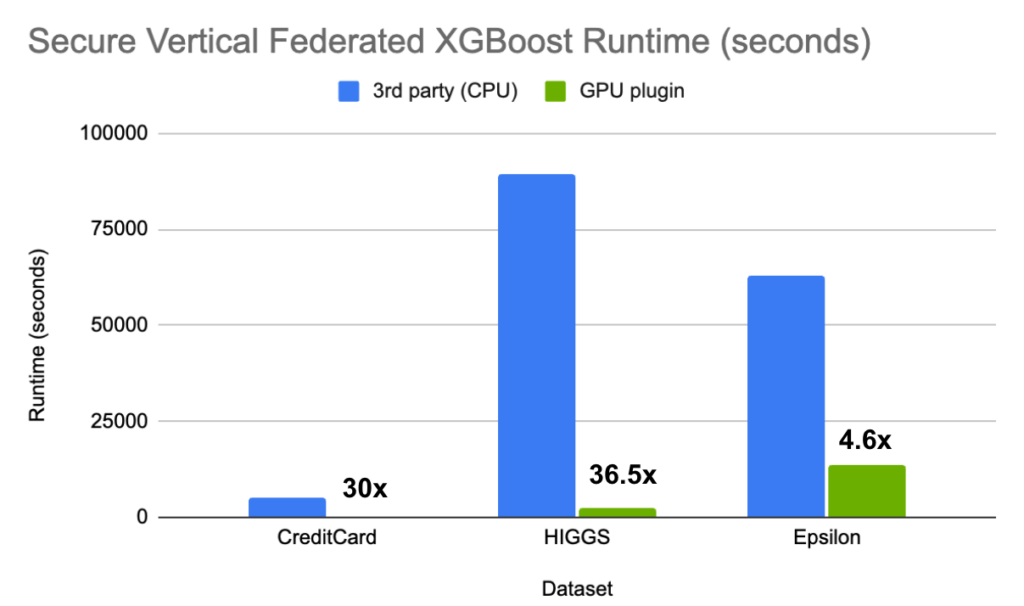}
\caption{Speed comparisons by different HE solutions for secure vertical federated XGBoost.}
\label{fig:3}
\end{figure}

\begin{figure}[htbp]
\centering
\includegraphics[width=0.45\textwidth]{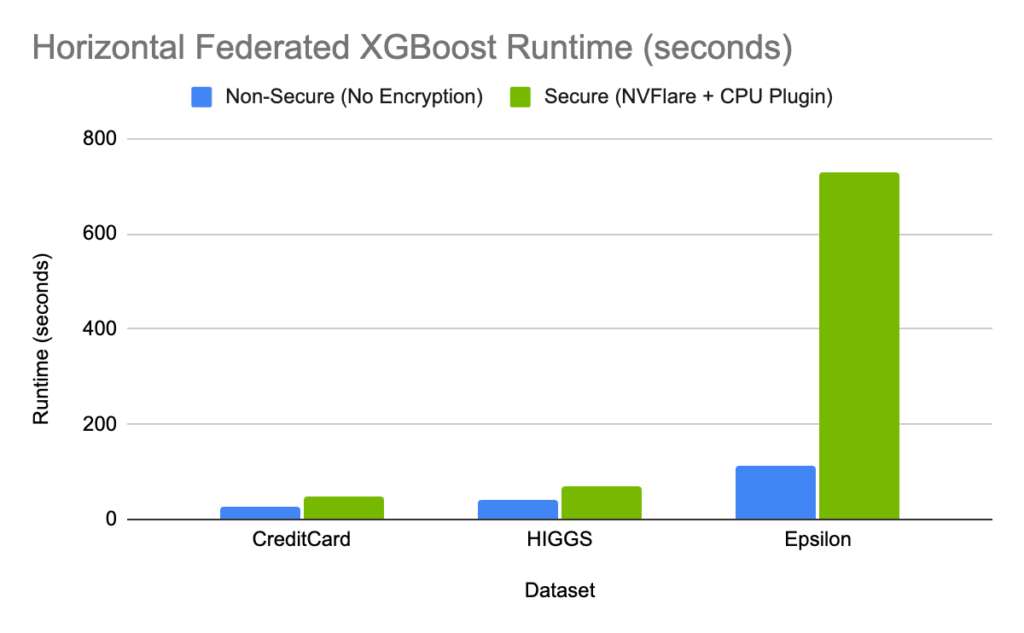}
\caption{Runtime of secure versus non-secure horizontal Federated XGBoost.}
\label{fig:4}
\end{figure}

End-to-end XGBoost training was performed with the following parameters: $num_{trees} = 10$, $max_{depth} = 5$, $max_{bin} = 256$. Testing was performed using the NVIDIA Tesla V100 GPU and the Intel E5-2698 v4 CPU. Fig.s~\ref{fig:3} and~\ref{fig:4} show the time comparisons. Note that the simulation was run on the same machine, so federated communication cost is negligible. 

For secure vertical federated XGBoost, we compare the time cost of the NVIDIA Flare pipeline CUDA-accelerated Paillier plugin (noted as GPU plugin) with the existing third-party open-source solution for secure vertical federated XGBoost. Both are HE-encrypted. Fig.~\ref{fig:3} shows that our solution is 4.6x to 36x faster depending on the combination of data and feature sizes. Note that the third-party solution only supports CPU.

For secure horizontal federated XGBoost, third-party offerings do not have a secure solution with HE. Therefore, we compare the time cost of the NVIDIA Flare pipeline without encryption and with the encryption plugin of CKKS using CPU (noted as the CPU plugin) to get an idea of the overhead of the encryption for data protection. 

As shown in Fig.~\ref{fig:4}, in this case the computation is notably faster than in the vertical scenario (orders of magnitude lower), and thus GPU acceleration may not be required with such reasonable overhead. Only for datasets with very wide histograms (Epsilon, for example), the encryption overhead will be more significant (but still approximate only 5\% of the vertical setting).   

\section*{Conclusion}
In this work, we demonstrated how GPU-accelerated Homomorphic Encryption enhances the security of Federated XGBoost, enabling privacy-preserving horizontal and vertical federated learning through NVIDIA FLARE. As compared with existing works of federated XGBoost, the new functionality provides 1) a secure federated XGBoost pipeline ensuring data safety from the algorithm level, and 2) an efficient CUDA-accelerated solution that is much faster than current alternatives on the market enabled by GPU computation. This will encourage adaptations in the fields that have high requirements over both data security and learning efficiency, where XGBoost is commonly used, such as fraud detection model training in the financial industry.   
 
Future research could further focus on optimizing finer-grained encryption selections according to the characteristics of the dataset with regard to sample number and feature number, allowing for improved efficiency. 

\bibliographystyle{IEEEtran}
\bibliography{secure_xgb}
\end{document}